\documentclass[namedreferences]{SolarPhysics}
\usepackage[optionalrh]{spr-sola-addons} % For Solar Physics 
\usepackage{graphicx}        % For eps figures, newer & more powerfull
\usepackage{color}           % For color text: \color command
\usepackage{url}             % For breaking URLs easily trough lines
\newcommand{\mres}{multiresolution}
\newcommand{\Mres}{Multiresolution}
\newcommand{\ls}{length-scale}

\newcommand{\bgdcl}{$\beta\gamma\delta$}
\newcommand{\bgcl}{$\beta\gamma$}
\newcommand{\bcl}{$\beta$}
\newcommand{\acl}{$\alpha$}
\newcommand{\linename}{MOPRS}

\begin{document}
\begin{article}
\begin{opening}

  \title{\Mres\ analysis of active region magnetic structure and its
    correlation with the Mount Wilson classification and flaring
    activity}

\author{J.     \surname{Ireland}$^{1}$,
        C.A.   \surname{Young}$^{1}$,
        R.T.J. \surname{McAteer}$^{2}$,
        C.     \surname{Whelan}$^{3}$,
        R.J.   \surname{Hewett}$^{4}$,
        P.T.   \surname{Gallagher}$^{5}$}

\runningauthor{Ireland et al.}
\runningtitle{\Mres\ analysis of active region magnetic structure}

\institute{ $^{1}$ ADNET Systems, Inc., NASA's Goddard Spaceflight
  Center, Mail Code 671.1, Greenbelt, MD 20771, USA.\\
  $^{2}$ Catholic University of America, NASA Goddard Space Flight
  Center, Greenbelt, MD 20771, USA.\\
  $^{3}$ School of Physics, University Col lege Dublin, Belfield,
  Dublin 4, Ireland.\\
  $^{4}$ Computer Science Departement, University of Il linois at
  Urbana-Champaign, Urbana, IL 61801, USA.\\
  $^{5}$ Astrophysics Research Group, School of Physics, Trinity 
  College Dublin, Dublin 2, Ireland.\\
}

\begin{abstract}
  Two different \mres\ analyses are used to decompose the structure of
  active region magnetic flux into concentrations of different size
  scales.  Lines separating these opposite polarity regions of flux at
  each size scale are found.  These lines are used as a mask on a map
  of the magnetic field gradient to sample the local gradient between
  opposite polarity regions of given scale sizes.  It is shown that
  the maximum, average and standard deviation of the magnetic flux
  gradient for $\alpha, \beta, \beta\gamma$ and $\beta\gamma\delta$
  active regions increase in the order listed, and that the order is
  maintained over all \ls s. Since magnetic flux gradient is strongly
  linked to active region activity, such as flares, this study
  demonstrates that, on average, the Mt. Wilson classification encodes
  the notion of activity over all \ls s in the active region, and not
  just those \ls s at which the strongest flux gradients are found.
  Further, it is also shown that the average gradients in the field,
  and the average length-scale at which they occur, also increase in
  the same order.  Finally, there are significant differences in the
  gradient distribution, between flaring and non-flaring active
  regions, which are maintained over all \ls.  It is also shown that
  the average gradient content of active regions that have large
  flares (GOES class 'M' and above) is larger than that for active
  regions containing flares of all flare sizes; this difference is
  also maintained at all \ls.  All the reported results are
  independent of the \mres\ transform used.  The implications for the
  Mt. Wilson classification of active regions in relation to the
  \mres\ gradient content and flaring activity are discussed.
\end{abstract}
\keywords{Sun: active region, Sun: magnetic field}
\end{opening}

\section{Introduction}
\label{Introduction} 

Many attempts have been made to assess the complexity of active
regions, as a measure of their activity.  The earliest attempt (dating
from 1908) - Mount Wilson classification - and still in use today,
puts active regions into four broad classes, and is based on the
distribution of magnetic flux polarity in the active region.  The
magnetic flux distribution, in relation to the location of white
light sunspots, in conjunction with some simple classification rules,
are used by the observer, to classify the observed region.  The four
main classes of active region, and their classification rules are
shown in Table \ref{tab:mtwilson}.

The success of the Mt. Wilson classification scheme lies in the fact
that it is simple, and has some predictive power when combined with
flare frequency rates deduced from long time series (many solar
cycles) observations. Other classification schemes exist, notably the
McIntosh classification scheme \cite{1990SoPh..125..251M}.  This is
a significantly more complex classification scheme because it looks at
the magnetic structure of the active region in much greater detail.
This too has some predictive ability in divining activity, again based
on correlation between flare frequency rates and the the McIntosh
class.

Both schemes are implemented manually, and so are subject to the usual
human observer biasses.  Research is currently ongoing into automating
these classification schemes, with some success
(\opencite{2006SoPh..239..519D}; \opencite{2002ApJ...568..396T};
\opencite{1998A.AS..131..371B}).  The notable feature about both the
Mt. Wilson and McIntosh classification schemes is that they both use
the magnetic structure, that is, the relative locations and sizes of
concentrations of opposite polarity magnetic flux in the active region
in order to determine active region class.  Since opposite polarity
flux creates gradients, these classification schemes are in
essence classifying the location and size of the magnetic flux
gradients in the magnetic data.  Hence these classification schemes
act as proxies for the magnetic flux gradient information content of
the active region.

The notion of complexity is ill-defined, and that carries over to the
study of active regions.  Many studies have attempted to quantify the
notion of active region complexity. \opencite{2005ApJ...631..628M}
look at around 10,000 active regions and calculate a fractal dimension
for the spatial distribution of the absolute magnetic flux.  Separated
by Mt. Wilson class, active regions are surprisingly homogenous in
fractal dimension, with very little difference between $\alpha$ class
active regions and $\beta\gamma\delta$ active regions.  This shows,
under the assumption that the flux is a mono-fractal, that all the
magnetic flux is basically the same, regardless of polarity and of
what class of active region it appeared in.

The mono-fractal restriction can be removed by assuming that the
magnetic flux can be described my a multifractal.  This assumes that
the flux can be represented by a distribution of fractals, a
considerable increase in sophistication.  This is physically
motivated, since the multifractals have deep ties to turbulence (
\opencite{2005SoPh..228....5G}; \opencite{2005SoPh..228...29A};
\opencite{2002ApJ...577..487A} ).  \opencite{conlon} present the
results of analyzing the multifractal content of several active
regions.  It is shown that the multifractal spectrum changes with the
emergence of the flux, indicating that the scale content is changing
with emergence.  This study allows for flux cancellation in
calculating the multifractal spectrum, and so implicitly takes account
of the different flux polarity, a significant difference from the
study of \opencite{2005ApJ...631..628M}.

Both these studies are more concerned with the nature of the flux, as
opposed to its gross distribution on the Sun's surface, which is what
the classification schemes outlined above deal with. In particular,
these studies do not make any statements about the gradient between
flux elements, which is known to be a significant indicator of
activity. This study is complementary, in that it attempts to look at
the spatial distribution of flux at different sizes, or \ls s, as
opposed to the nature of that flux.  To do this, the study uses four
\mres\ analyses to decompose the structure on different \ls s, and
then looks at the gradient between flux elements of different size in
an attempt to examine how the scale size of the spatial flux
distribution is related to the gradient, and to the Mt. Wilson
classification, which implicitly encodes gradient information.  It is
known that the presence of strong gradients in the magnetix flux
distribution is indicative of the active region activity
(\opencite{2007ApJ...656.1173L}; \opencite{2006ApJ...644.1258F};
\opencite{2003ApJ...595.1277L}(a); \opencite{2003ApJ...595.1296L}(b);
\opencite{2002ApJ...569.1016F}; \opencite{1997ApJ...482..519F}).  A
recently found indicator of activity is given by
\opencite{2007ApJ...655L.117S}, in which it is found that the total
unsigned magnetic flux in a 15 arcsecond strip around strong polarity
inversion lines is an excellent predictor for the occurence of M and X
class flares within 24 hours of observation.  All these studies
implicate gradients in the magnetic field as being a crucial component
indicating the likelihood of activity (\opencite{cui};
\opencite{2002SoPh..209..171G}).

Section \ref{sec:multi} describes the \mres\ analyses used.  Section
\ref{sec:data} describes the data used in this study.  Section
\ref{sec:results} describes and discusses the results of the analysis
procedure.

\begin{table}
\begin{tabular}{cl}
class         & feature/classification rule \\
$\alpha$      & a single dominant spot often linked with a plage of opposite magnetic polarity\\
$\beta$       & a pair of dominant spots of opposite polarity\\
$\gamma$      & complex groups with irregular distribution of polarities\\
$\beta\gamma$ & bipolar groups which have more than one clear north-south polarity inversion line\\
$\delta$      & umbrae of opposite polarity together in a single penumbra\\
\end{tabular}
\caption{Mt. Wilson classification rules}
\label{tab:mtwilson}
\end{table}

\section{Magnetic flux data}
\label{sec:data}
The magnetic flux data used in this paper is the same as that used by
(\opencite{2005ApJ...631..628M}; also \opencite{2005SoPh..228...55M})
in their study of active region fractal dimension.  The analyzed
dataset is based on Solar and Heliospheric Observatory (SoHO)
Michelson Doppler Imager (MDI) images (\opencite{1995SoPh..162....1D};
\opencite{1995SoPh..162..129S}, and consists of extracted subfields
from full disk MDI magnetograms, centred on active regions present on
the disk. Full details of the method of extraction and correction can
be found in \opencite{2005ApJ...631..628M}.  The final dataset for use
in the present study consists of 19827 $600''\times600''$ FITS
(Flexible Image Transport System) files each centred on one or more
active regions.  Of these, only those within 60 degrees of disk centre
have their magnetic structure decomposed using the algorithm described
in Section \ref{sec:algorithm}.  Images more than 60 degrees away from
disk centre contain too many artifacts from projection effects and
field reversals for the \mres\ analyses to proceed safely.  This
leaves 9757 usable active region images.

\section{\Mres\ algorithms}
\label{sec:multi}
Active regions are complex objects, and to attempt to understand the
spatial distribution of their flux sources, any analysis must look at
all the \ls s available.  This arises from the observation
that active regions emerge and occur in many different shapes, sizes,
and configurations, and the fact that activity is often confined to
very small localized regions.  This points naturally to a
\mres\ approach to understanding the flux distribution.  However,
since active regions are complex objects, it is inevitable that a
single analysis will not capture all the information of interest, and
so, for comparative purposes, four \mres\ algorithms are implemented,
enabling cross-checking of results.

Two simple algorithms are chosen, a wavelet transforms (Mexican hat)
and one \mres\ morphological transform (\mres\ median transform).
These mother wavelets are chosen since they apepar briadly similar to
the features we are attempting to isolate.  However, the wavelet-based
algorithm does lead to the introduction of spurious features in the
transform (which has the possibility of misleading further analysis,
see Section \ref{sec:algorithm}) and hence a second, completely
different analysis algorithm (which does not create these features)
based on the median filter, is also used.

\subsection{Wavelet transforms}
\label{sec:mexican}
\begin{figure}
  \centerline{\hspace*{0.015\textwidth}
    \includegraphics[width=0.515\textwidth,clip=]{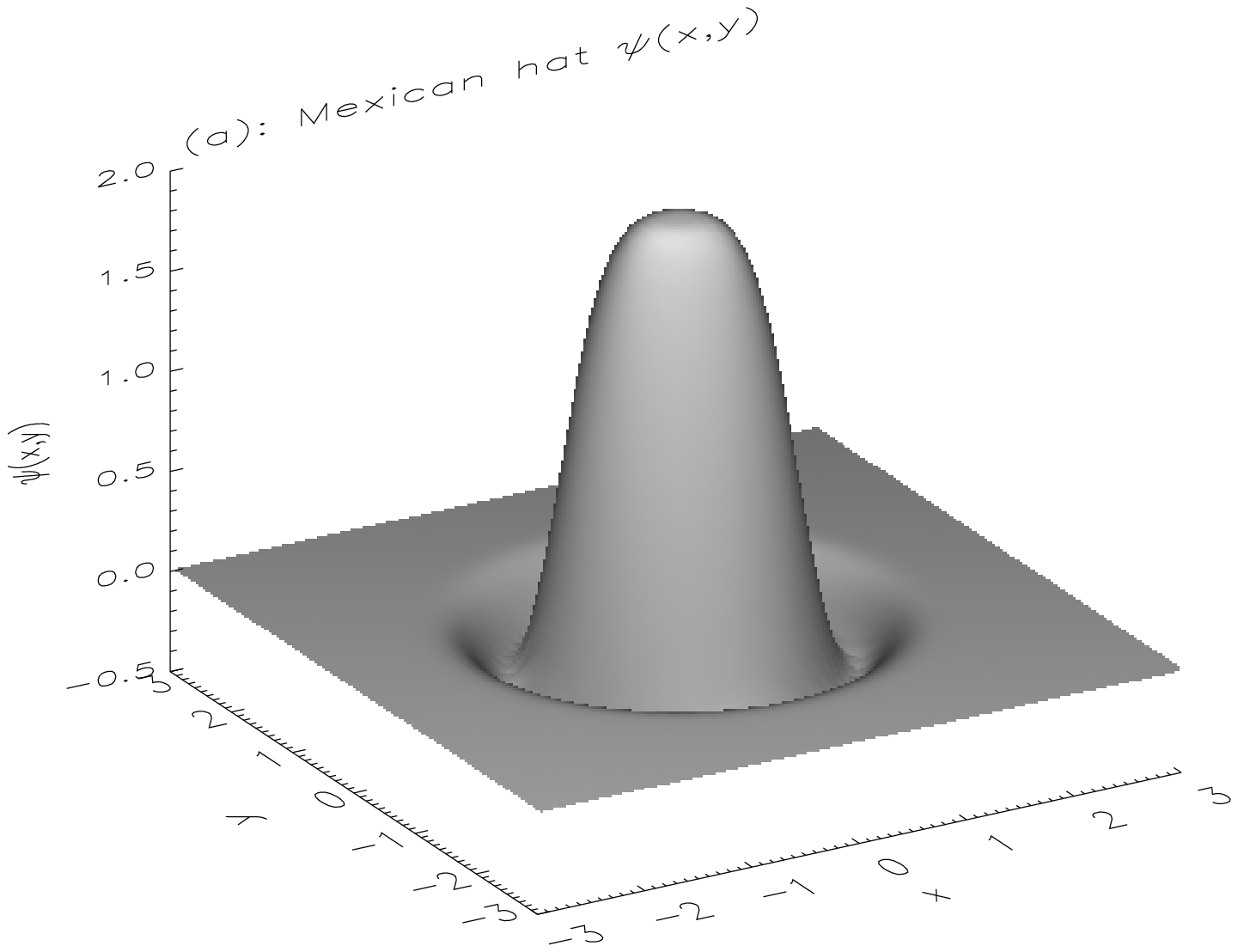}
    \hspace*{-0.03\textwidth}
    \includegraphics[width=0.515\textwidth,clip=]{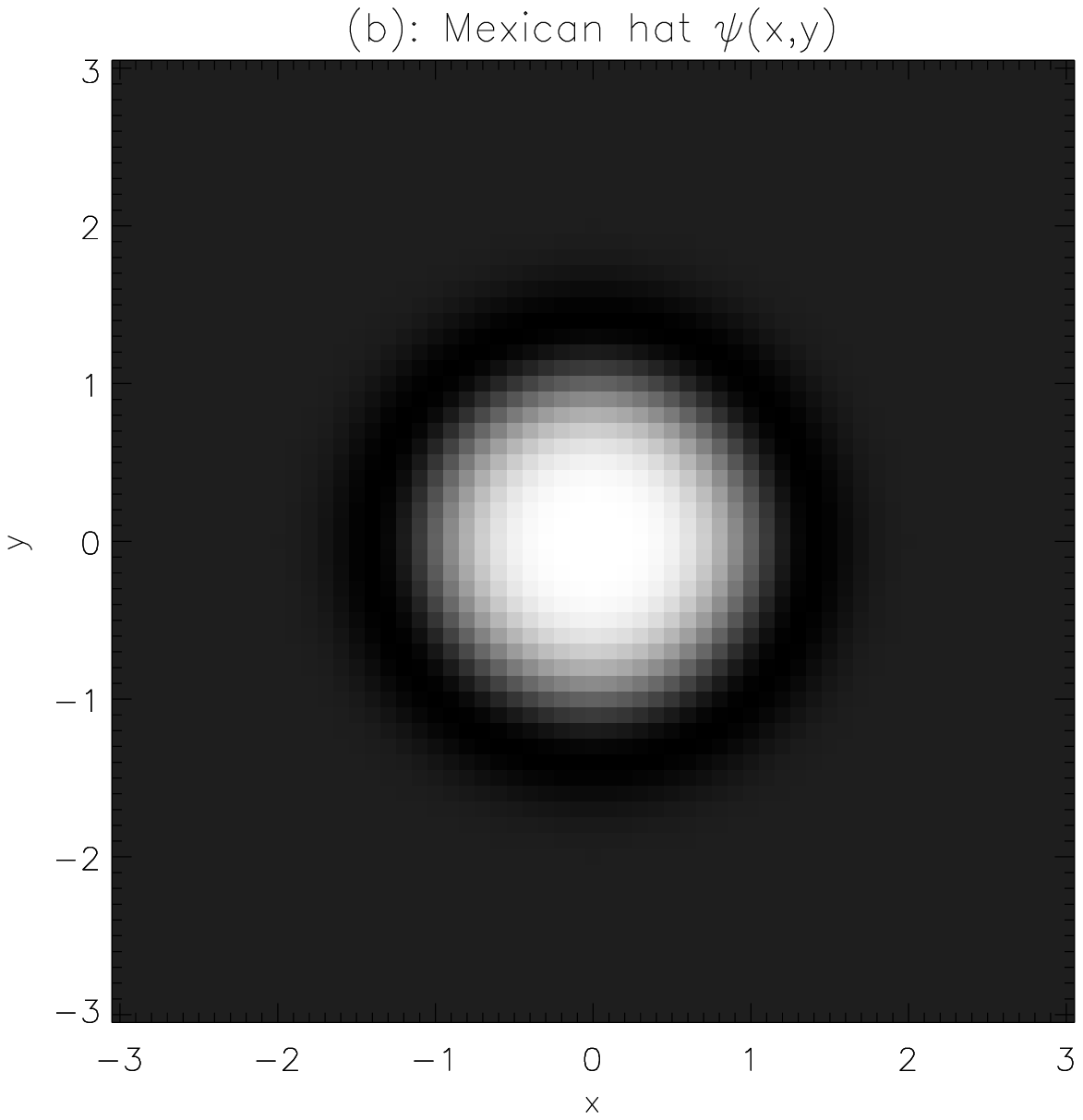}
  }
  \caption{Mexican hat mother wavelet, as a (a) surface and (b) contour
    plot. Interpreting the greyscale as opposite polarity flux, the
    single Mexican hat central flux concentration has a ring of opposite
    polarity flux.  These features are also evident in the transform at
    all \ls s. }
  \label{fig:mexhat}
\end{figure}

The continuous wavelet transform of a two-dimensional image $S$ in the
domain $D\subset R^{2}$ is
\begin{equation}
W(L,\underline{x}') = \int_{D}
\frac{1}{L}\psi
\left(
\frac{\underline{x}-\underline{x}'}{L}
\right)
S(\underline{x}) d\underline{x}
\end{equation}
where $L$ is the wavelet scale and $\underline{x}'$ translates
the wavelet across the domain $D$.  The Mexican hat wavelet transform
is continuous, and is implemented via the mother wavelet
\begin{equation}
\psi(\underline{x}) = c(2-|\underline{x}|^2)\exp\left(-|\underline{x}|^{2}/2 \right)
\label{eqn:mother}
\end{equation}
for some normalization constant $c$.  A plot of this mother wavelet is
shown in Figure \ref{fig:mexhat} (this wavelet is also used in the
analysis of \opencite{hewett}).  The wavelet is isotropic,
non-orthogonal, and is a good approximation to the shapes found in
active region magnetic fields.

\subsection{\Mres\ median transform}
\label{sec:median}
As noted above, the wavelet transforms used above are good at
identifying point and approximately circular features in images, but
have the unfortunate side effect of introducing a ``ringing'' which
appears as a false opposite polarity flux in the transform (Figure
\ref{fig:mexhat}, Figure \ref{fig:ex:trans}(I:c-h); the same effect
is also clearly visible in Hewett et al. 2007).  It is clearly
desirable to have a transform in which positive (negative) structure
in the image does not create negative (positive) structure in the
transform.  One such transform which has this property is the \mres\
median transform, based on the median filter.

The median transform for an image $S$ at \ls\ $L$ (denoted by
$\mbox{med}(S,L)$) is found by sliding a kernel of dimensions $L\times
L$ at all points in the image and calculating the median value\footnote{
If the list $X_{1},...,X_{N}$ is ordered such that $X_{j}\le X_{j+1}, \forall j, 
1\le j \le N-1$, then \cite{2003psa..book.....W},
\[
\mbox{median}(X_{1},...,X_{N}) = \left\{
\begin{array}{cc}
X_{j} & \mbox{, $j=N/2 + 0.5$ if $N$ is odd}\\
(X_{j} + X_{j+1})/2 & \mbox{, $j=N/2$ if $N$ is even}.\\
\end{array}
\right.
\label{eqn:med}
\]
} of the $L\times L$ subimage extracted from $S$.  The \mres\
median transform on $N_{l}$ (specified by the user) \ls s is
found by through the following algorithm:
\begin{enumerate}
\item  Let $c_{j} = S$ with $j = 1$. 
\item Determine $c_{j +1} = \mbox{med}( S , 2l + 1 )$. 
\item The \mres\ coefficients $w_{j+1}$ are defined as: $w_{j+1} = c_{j}-c_{j +1}$ . 
\item Let $j \leftarrow j + 1$; $l\leftarrow 2l$. Return to step 2 if $j < N_{l}$ . 
\end{enumerate}
The original image $S$ may be reconstructed by $S = c_{N_{l}}
+\sum_{j=1}^{N_{l}}w_{j}$ where $c_{N_{l}}$ is the residual image left
after exiting the algorithm.  In step 4, the set of resolution levels
associated with $S$ lead to a dyadic decomposition.  A useful
feature of the median transform for analysis of active regions is that
the shapes of structures on the analyzed \ls s are closer to
those of the input image than would be case with a wavelet transform.
The \mres\ median transform acts like a median filter operating on
multiple length-scales.  \cite{gonzaleswoods} state that the median
filter forces neighboring pixels to become more like their neighbors.
This operation will preserve the structure of the field, whilst
ignoring noisy outliers that may skew wavelet based processing.

\subsection{An algorithm to define a Multi-scale opposite polarity region separator (\linename)}
\label{sec:algorithm}
\begin{figure}
\begin{tabbing}
\bf{Algorithm ``\linename''} \\
1 be\=gin   \\
2   \> R\= emove noise in image $S$ thresholding: pixels where $|B|<50$\\
    \>  \> are set to zero.   \\
3   \> D\= ecompose $S$ by \mres\ transform onto $n$ \ls s, $L_{i}$ \\
    \>  \> to get transforms $W_{i}$, $1\le i \le n$. \\
4   \> for \= i in range 1,n  \\
5   \>          \> Find zero contours in $W_{i}$ \\
6   \>          \> K\= eep pixel $p_{i,a,b}$ on the zero contour if there exists opposite \\
    \>          \>  \> polarity fields within a 3x3 box centred on that pixel.\\
    \>          \>  \> in the original image $S$. \\
7   \>          \> Report the locations of all the lines at found at this \ls -  $M_{i}$ \\
8   \>          \> Calculate the local transform power around each line in $M_{i}$.    \\
9    \> end \\
10 end \\
\end{tabbing}
\caption{Algorithm to find Multi-scale Opposite Polarity Region
  Separators (\linename).}\protect\label{fig:algorithm}
\end{figure}

\begin{figure}
  \centerline{
    \hspace*{-0.015\textwidth}
    \includegraphics[width=0.98\textwidth,height=0.25\textheight,clip=]{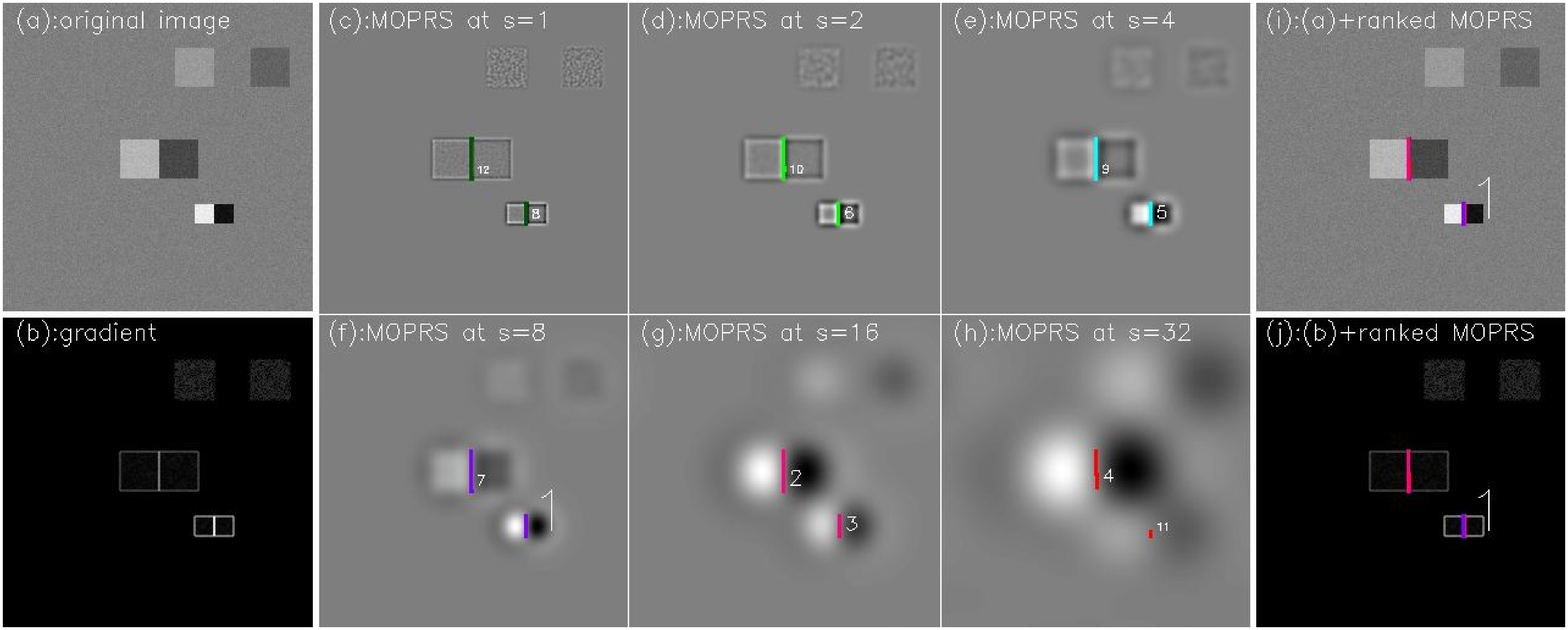}
  }    
  \vspace*{0.005\textheight}
  \centerline{
    \hspace*{-0.015\textwidth}
    \includegraphics[width=0.98\textwidth,height=0.25\textheight,clip=]{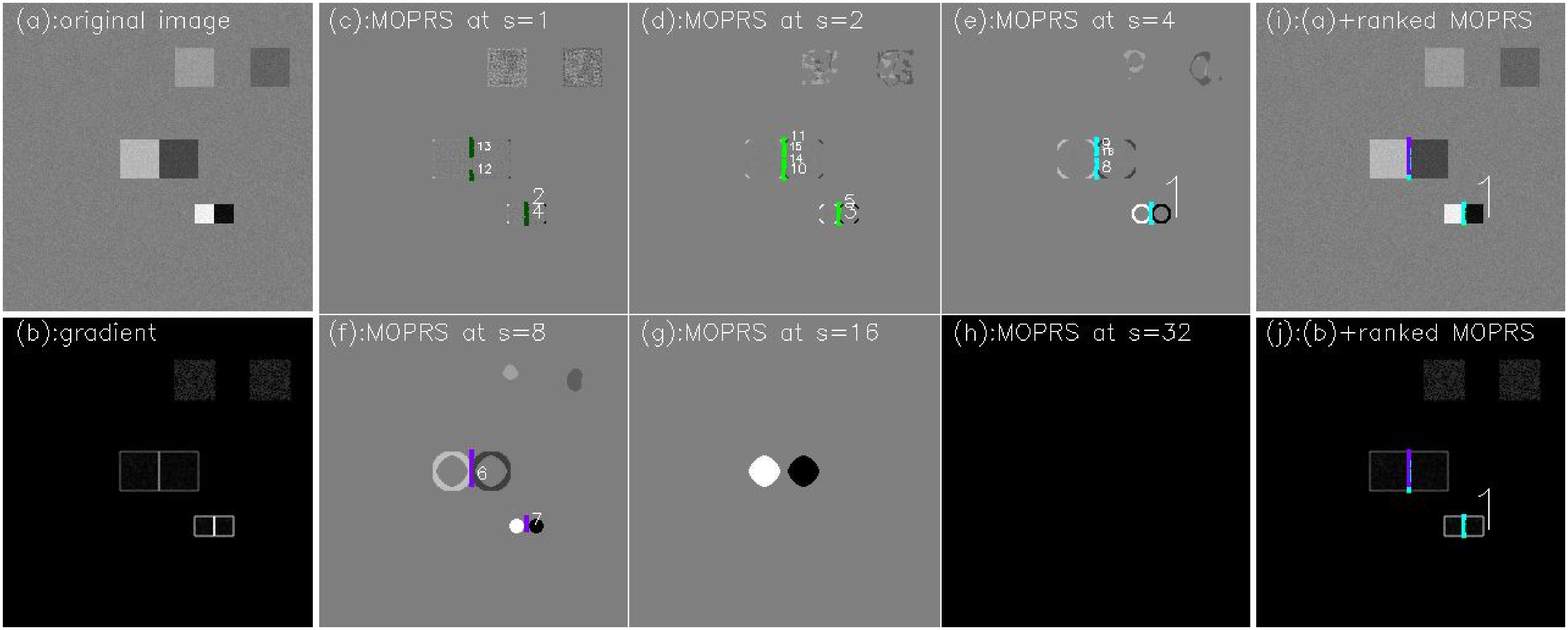}
  }
  \vspace{-0.80\textwidth}   % Shift close to the panel top 
  \centerline{\large \bf     % Includes the labels (here needs the color package)
    \hspace{0.0 \textwidth} \color{white}{(I)}
    \hfill}
  \vspace{0.375\textwidth}    % Shift back to the panel bottom 
  \centerline{\large \bf     % Includes the labels (here needs the color package)
    \hspace{0.0 \textwidth} \color{white}{(II)}
    \hfill}
  \vspace{0.35\textwidth}    % Shift back to the panel bottom 
  \caption{ Panels (I): an example \linename\ analysis using the Morlet
    wavelet (Section \ref{sec:mexican}). Panel I(a) shows some test
    data, I(b) the magnitude of the gradient, and I(i), I(j) show
    I(a), I(b), respectively with \linename\ overlaid.  Panels I(c-h) show
    the wavelet transform of test data at length-scale $s$, with the
    \linename\ found at that length-scale.  Each line is labeled with its
    local wavelet power rank.  Panels (ii): the same data is analyzed
    by the median transform.  Panels II(a,b,i,j) have the same meaning
    as panels I(a,b,i,j). Panels II(c-h) show the median transform of
    the test data at length-scale $s$, with the \linename\ found at that
    length-scale.  Each line is labeled with its local median
    transform power rank.}
  \label{fig:ex:trans}
\end{figure}

At a given \ls, the wavelet transforms decompose the active region
into objects of the same \ls\ (Figures \ref{fig:ex:trans}(I:a-j).  The
\mres\ median transform is slightly different in that the active
region field structure is not directly compared to a given shape;
rather, the \mres\ coefficients arise from the distribution of
magnitudes of active region fluxes in the analysis window (note
however that the shape of window used to define the local data at that
\ls\ is square.)

The organization of the active region at different scale sizes and
different polarities is apparent.  This makes it possible to find
lines slicing the active region structure between opposing polarity
regions of a given scale size.  These lines are termed multi-scale
opposite polarity region separators (\linename), and they delineate
the organization of opposite polarity regions in the active region.
The algorithm describing the generation of \linename\ is given in
Figure \ref{fig:algorithm}.  These lines are geometrical constructions
based on the \mres\ analysis used, and since the transforms we are
using are different, we can expect the \linename\ defined to vary.

The \linename\ are used to {\it sample} the gradient of the magnetic flux
between regions of opposite polarity flux, at multiple \ls, using the
following algorithm.
\begin{enumerate}
\item Rank all the \linename-lines by local transform power,
  regardless of scale.
\item Create a mask by overlaying lines with larger local transform
  power over lines with smaller local transform power.  This enables
  overlapping \linename\ with stronger local wavelet power to be
  preferentially represented, since \linename\ with weaker local
  wavelet power are replaced.  Each \linename\ carries a label with the
  \ls\ it was found at.
\item Return gradient of original image $S$ at each point on the remaining
  \linename, and assign that gradient to the \linename's \ls\ label.
\end{enumerate}
Regardless of which \mres\ transform is used, the \linename\ exist to
sample the same underlying magnetic field gradients, and so each
\mres\ transform is analyzing the same complex, physical object.
Different \mres\ transforms will give different samples of the same
underlying magnetic field gradient, and the results obtained have to
be interpreted with respect to the properties of the transform in
mind.  However, this is the case for all decompositions.  For example,
one can decompose any one dimensional time series signal using any
orthogonal set; using Hermite polynomials is mathematically equivalent
to using the more familiar sinusoids, and the scale content of the
signal can be assessed.  However, the advantage of sinusoids is that
they are also the normal modes of vibration for a number of different
systems, and therefore have an additional meaning on top of their
convenience in understanding the scale structure of a time series.  It
is possible to understand the scale structure of a one dimensional
time series with Hermite polynomials, but more difficult.
Unfortunately, there is little or no extra guidance in choosing which
\mres\ decomposition will be the ``best'' for understanding the
complex structure of active regions, other than looking for mother
wavelets that represent the object we are looking at.  Therefore, in
lieu of a ``best possible choice'', we instead use multiple \mres\
transforms.

Figure \ref{fig:ex:trans}(I) implements the Mexican hat transform and
\linename\ analysis for some test data.  Figures
\ref{fig:ex:trans}(I:a) shows the test data, whilst (I:b) shows the
gradient of the image.  Figures \ref{fig:ex:trans}(I) show the Mexican
hat wavelet transform of the original image at multiple \ls s.
Overplotted in color are the \linename\ found at that \ls, along with the
rank of each line.  These lines are then plotted on the original image
and the gradient image, Figures \ref{fig:ex:trans}(I:c,d).  Discussion
of the transformed images and the concomitant gradient distributions
are presented below.

\section{Results}
\label{sec:results}
The \mres\ analysis described above generates a large amount of
information for each active region.  Gradient information found along
the \linename s measures the distribution of gradients in the
line-of-sight magnetic field as a function of the \ls\ of opposite
polarity regions in active regions.  Summary statistics of the
gradient distributions (after grouping these according to Mt. Wilson
class, flaring activity and \ls) are discussed below, for the 9757
active region observations described in Section \ref{sec:data}.

The discussion begins with examining the behaviors of the transforms
themselves.

%
% alpha
%
\begin{figure}
  \centerline{
    \hspace*{-0.015\textwidth}
    \includegraphics[width=0.98\textwidth,height=0.25\textheight,clip=]{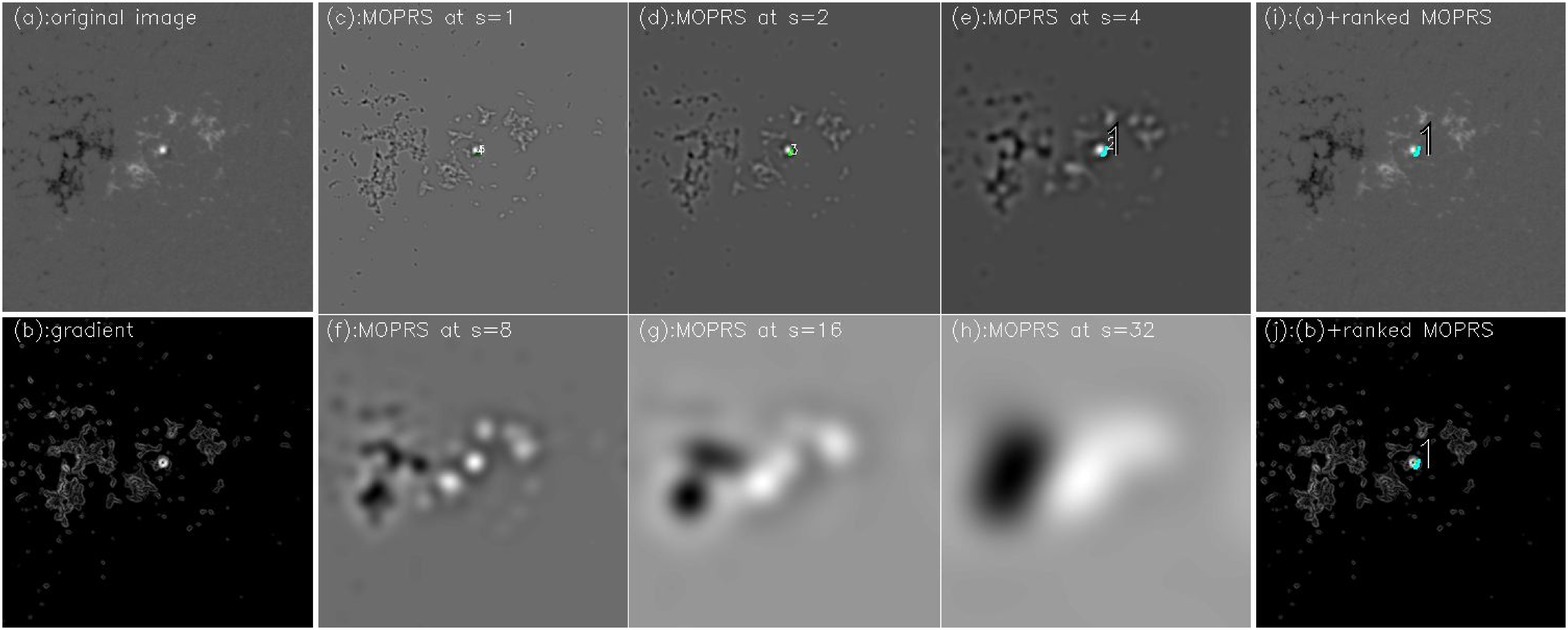}%{cor_mdi_ar7995_19961114_1251cor.fits.mex.all.eps}
  }    
  \vspace*{0.005\textheight}
  \centerline{
    \hspace*{-0.015\textwidth}
    \includegraphics[width=0.98\textwidth,height=0.25\textheight,clip=]{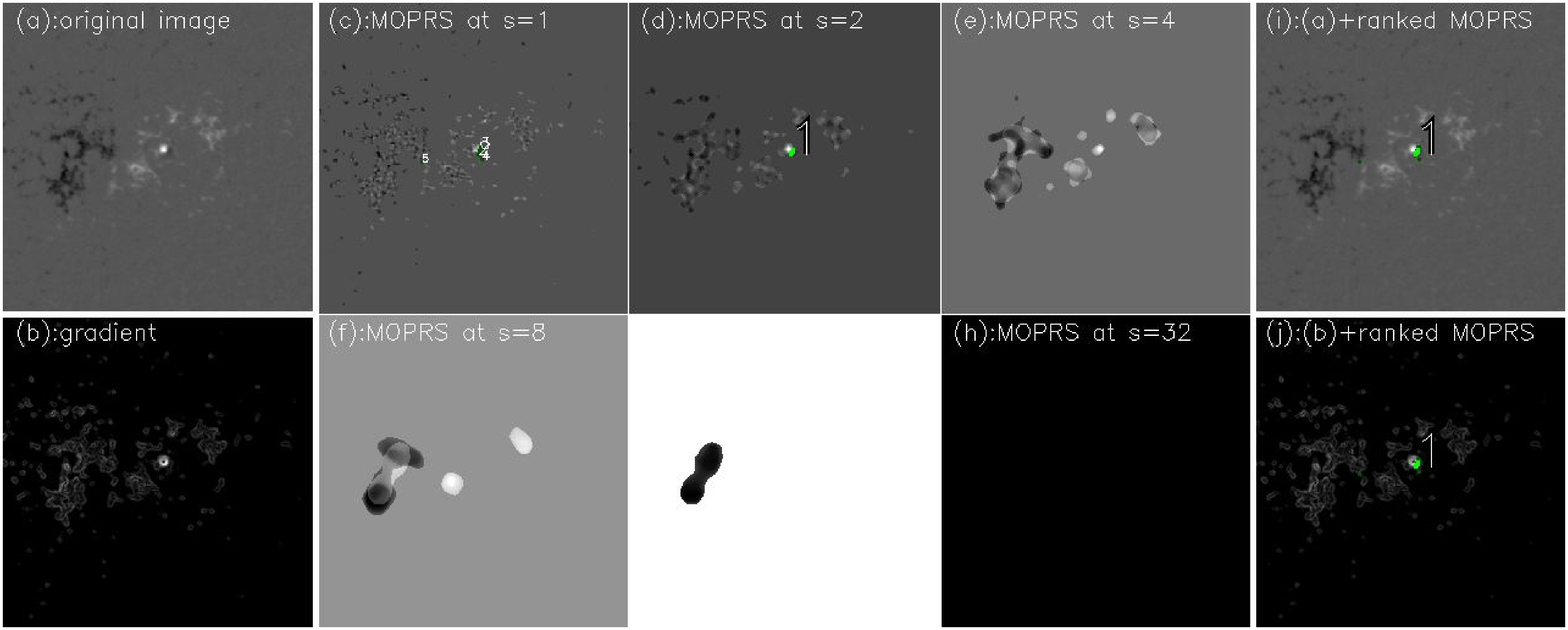}%{cor_mdi_ar7995_19961114_1251cor.fits.med.eps}
  }
  \vspace{-0.80\textwidth}   % Shift close to the panel top 
  \centerline{\large \bf     % Includes the labels (here needs the color package)
    \hspace{0.0 \textwidth} \color{white}{(I)}
    \hfill}
  \vspace{0.375\textwidth}    % Shift back to the panel bottom 
  \centerline{\large \bf     % Includes the labels (here needs the color package)
    \hspace{0.0 \textwidth} \color{white}{(II)}
    \hfill}
  \vspace{0.35\textwidth}    % Shift back to the panel bottom 
  \caption{Comparison of Mexican hat and median transform \linename\
    for an $\alpha$ region (NOAA AR 7995, 1996/11/14 12:51 UT). See
    Figure \ref{fig:ex:trans} for a description of the ordering of the
    plots.  A blank plot denotes that no information was found at that
    scale.  The region is very simple, and only a very few pixels are
    retained as \linename\ in the median transform compared to the
    Mexican hat results.}
  \label{fig:mexmed:a}
\end{figure}
%
% beta-gamma-delta
%
\begin{figure}
  \centerline{
    \hspace*{-0.015\textwidth}
    \includegraphics[width=0.98\textwidth,height=0.25\textheight,clip=]{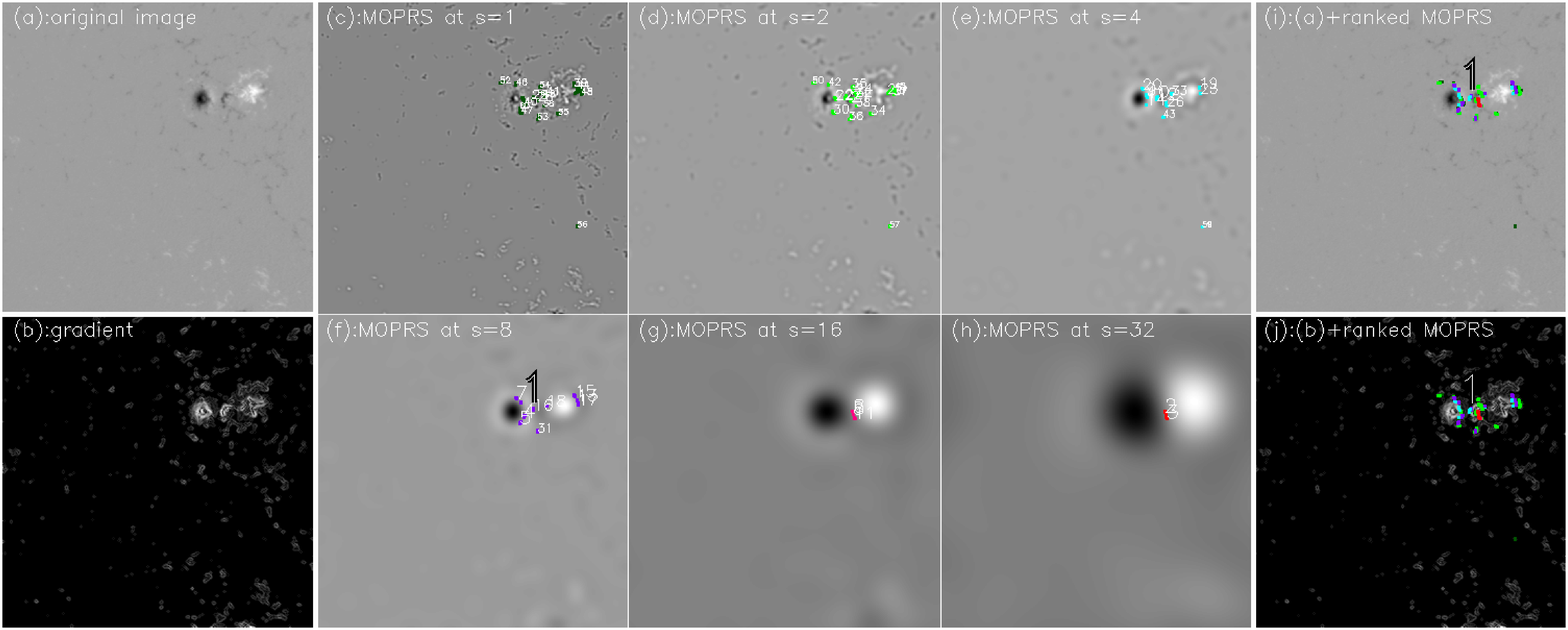}
  }    
  \vspace*{0.005\textheight}
  \centerline{
    \hspace*{-0.015\textwidth}
    \includegraphics[width=0.98\textwidth,height=0.25\textheight,clip=]{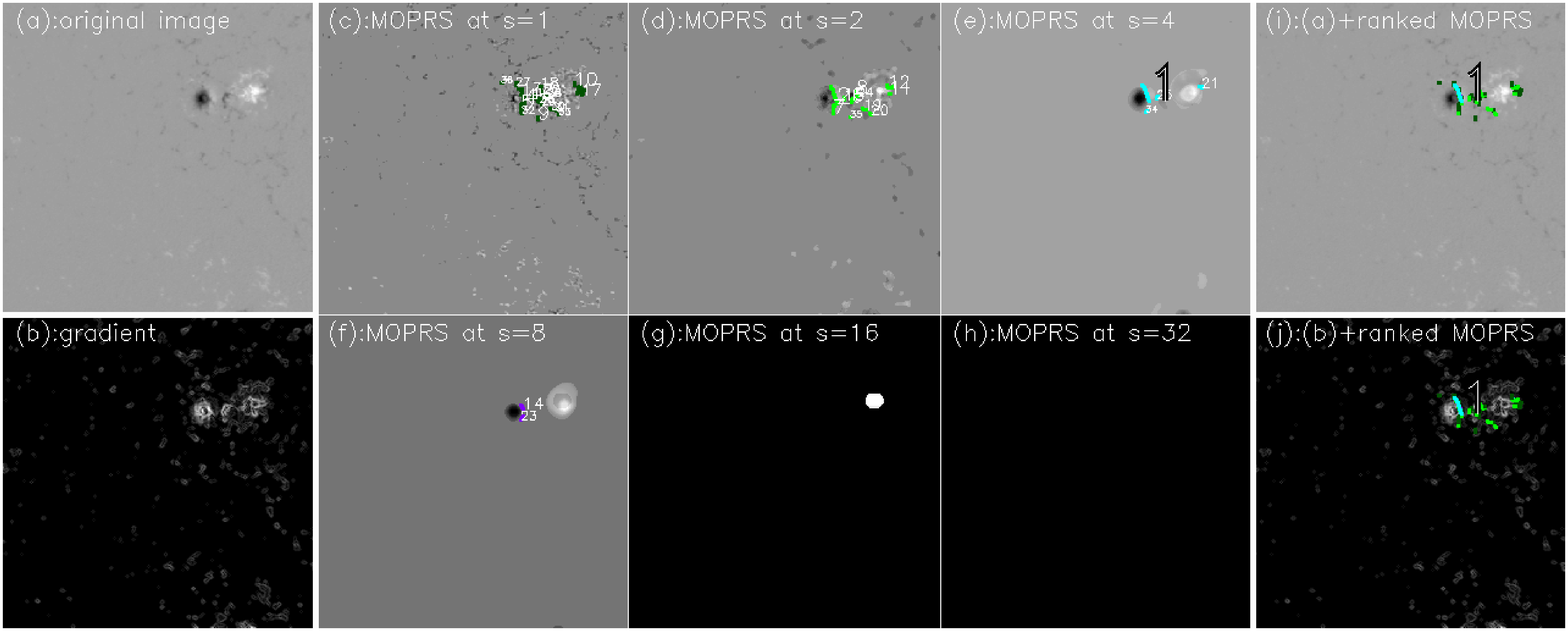}
  }
  \vspace{-0.80\textwidth}   % Shift close to the panel top 
  \centerline{\large \bf     % Includes the labels (here needs the color package)
    \hspace{0.0 \textwidth} \color{white}{(I)}
    \hfill}
  \vspace{0.375\textwidth}    % Shift back to the panel bottom 
  \centerline{\large \bf     % Includes the labels (here needs the color package)
    \hspace{0.0 \textwidth} \color{white}{(II)}
    \hfill}
  \vspace{0.35\textwidth}    % Shift back to the panel bottom 
  \caption{Comparison of Mexican hat and median transform \linename\
    for a $\beta\gamma\delta$ region (NOAA AR 2004/02/10 12:51 UT).
    The left two columns are Mexican Hat results, the right two
    columns are median transform results.  The individual plots for
    each type transform are (a) original magnetogram, (b) original
    magnetogram overplotted with \linename\ from all \ls s, (c-h)
    \mres\ transforms at \ls s L=1,2,4,8,16 and 32 pixels
    respectively, overplotted with \linename\ found at that scale.
    Note that the thickness of the \linename\ are weighted by the
    local magnetic flux gradient found at that location.  Both
    transforms capture the location of opposing polarities at small
    scales quite well.  The Mexican hat transform appears to slice the
    large scale structure of the region better than the median
    transform.}
  \label{fig:mexmed:bgd}
\end{figure}

\subsection{Comparative behaviour of the transforms}
\label{sec:compare}
It is apparent from the example transforms given in Figures
\ref{fig:ex:trans}, \ref{fig:mexmed:a} and \ref{fig:mexmed:bgd} that
each transform captures different information about the active region.
For example, at large scales, Figures \ref{fig:ex:trans}(I:g,h) and
\ref{fig:mexmed:a}(I:g,h) divide the plane into approximately two
regions of opposite polarity, capturing the large scale distribution
of the field (Figure \ref{fig:ex:trans}(I:a),
\ref{fig:mexmed:a}(I:a)). The median transform does not return any
structure at the largest scale size (Figure \ref{fig:ex:trans}(II:h),
\ref{fig:mexmed:a}(II:h)).  The ``ringing'' of the Mexican
hat transform is also apparent in Figures \ref{fig:ex:trans}(I:c-h),
brought about by the shape of the transform (Section
\ref{sec:mexican}).  Both \mres\ analyses report \linename\ where one
would expect them, but give them different rankings.  In particular,
the highest ranked \linename\ occurs at different (and neighboring \ls
s), but at the same location.

The median transform is much more localized, and so even although it
does also find \linename\ at larger \ls s, the pixels retained as
having significant gradients are far fewer.  This is due to the
interaction of the \linename\ algorithm (Figure \ref{fig:algorithm})
with the properties of the transform.  The median transform cleaves
more closely to the true shape of the regions
\cite{1998ipda.book.....S} at that \ls, whereas the Mexican hat
transform imposes a shape on the active region.  As very large scale
contiguous features are not very common in active regions, they are
not there for the median transform to find, and so \linename\ at these
\ls s are deprecated compared to the Mexican hat transform.  This is
apparent in Figure \ref{fig:mexmed:bgd}(II:g,h), where the large scale
organization of the field is found via the Mexican hat transform
because the transform at that scale is smoothing out the original data
over a large \ls.  This is analogous to the situation in Morlet
wavelet time series analysis, where a spike (delta function) in the
time series can leads to power at larger scales, even although there
is no real variation at those longer scales in the data
\cite{2004SoPh..222..203D}.  The median transform does not detect any
information at this \ls\ (Figure \ref{fig:mexmed:bgd}(II:g,h).

Line 5 in the \linename\ algorithm (Figure \ref{fig:algorithm}) looks
for zero contours in the transform, and so can be susceptible to the
ringing effect of the Mexican hat transform.  This effect is
suppressed by line 6 of the \linename\ algorithm, by referring back to
the original data to look for pixels on the \linename\ which do have
some opposite polarity flux around them.  The overall result is that
the Mexican hat transform is quite effective at dividing the active
region into intuitively satisfying regions, at a given \ls.  The
median transform is more effective at retaining the local shape of
opposite polarity regions at a given \ls, and hence more effective at
describing where opposite polarity flux is at close proximity. The
\linename\ ranking algorithm also reports the most ``important''
\linename\ at different \ls\ (Figures \ref{fig:ex:trans},
\ref{fig:mexmed:bgd}).  This is not very surprising, as the two \mres\
transforms are measuring very different properties of the structure.
In general, it is found that the median transform assigns \linename\
to smaller \ls\ than the Mexican hat transform, consistent with the
behavior noted by \opencite{1998ipda.book.....S}.  The results quoted
below from each method are consistent as a function of \ls, as defined
by each analysis method.  Since the results are qualitatively the same
for each method, this confirms that each method is indeed internally
consistent, and both are measuring the same property of the active
region field.

\subsection{Gradients along \linename}
\label{sec:res:rlm}
\begin{figure}
  \centerline{\hspace*{0.0\textwidth}
    \includegraphics[width=0.5\textwidth,viewport =  0 0 275 360, clip=]{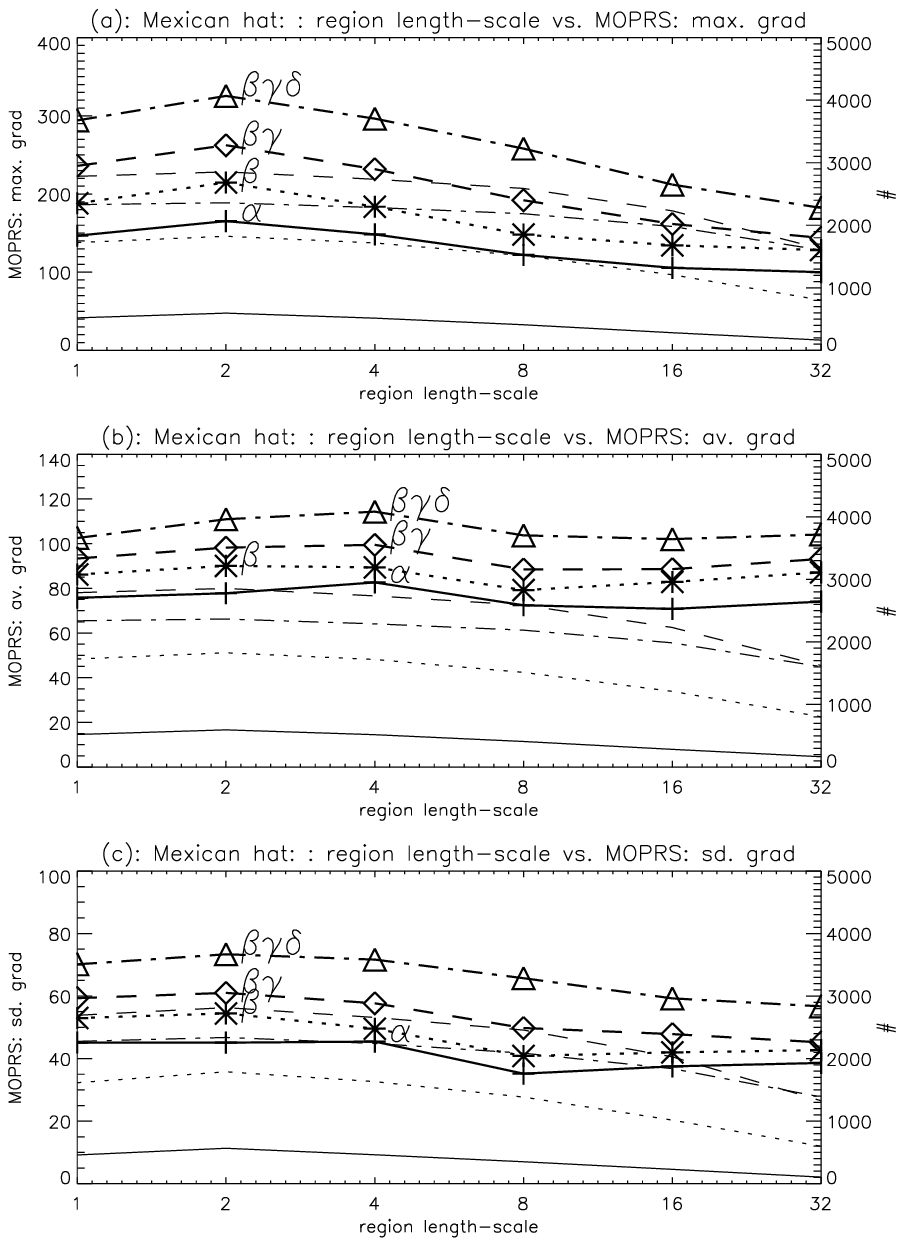}
    \hspace*{0.0\textwidth}
    \includegraphics[width=0.5\textwidth,viewport =  0 0 275 360, clip=]{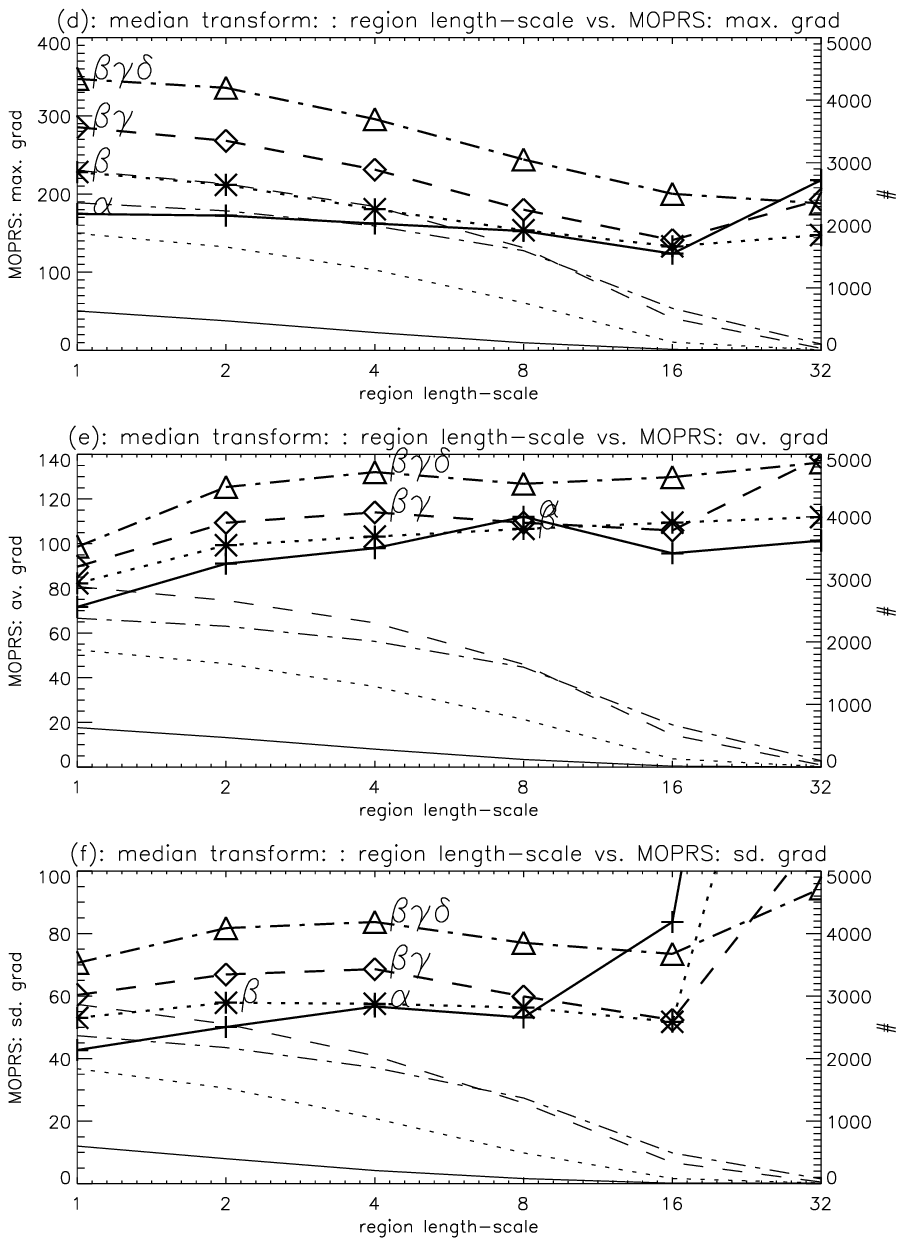}
  }
  \caption{Measurements of the gradient statistics as a function of
    multi-resolution transform (Mexican hat, a-c; median transform,
    d-f), Mt. Wilson classification ($\alpha$, solid line style;
    $\beta$, dotted; $\beta\gamma$, dashed; $\beta\gamma\delta$,
    dot- dashed) and \linename\ \ls (thick lines with plot
    symbols). Plots (a,d) show the average {\it maximum} gradient
    found on each \linename; plots (b,e) show the average {\it
      average} gradient on each \linename\ and plots (c,f) show the
    average {\it standard deviation} of the gradient found on each
    \linename.  Also indicated are the total number of images that
    have at least one \linename\ at any \ls\ and return a valid
    gradient measurement, as a function of Mt. Wilson classification
    (thin lines, same line style as the Mt. Wilson classification,
    numbers on left hand plot axis).
    These can vary per plot since occasionally the analysis will find
    a \linename\ that will return a gradient statistic which is
    undefined: for example, occasionally the \linename\ consists of a
    single point - in such a case the standard deviation of the
    gradient underlying this point is undefined and so must be
    excluded from further analysis.  See Section \ref{sec:res:rlm}
    for more discussion of this result.}
\label{fig:res:rlm}
\end{figure}
Figure \ref{fig:res:rlm}(a-c,d-f) shows the results of
measuring the gradients along the \linename\ for the active region
data set for the Mexican hat and median transforms respectively.  Four
different statistics are returned; (a,d), maximum gradient found along
the \linename\, (b,e), average gradient found along the \linename\,
(c,f), standard deviation of the gradient along the \linename\.  In
all cases, the average value of the quantity is plotted as a function
of \ls\ and Mt. Wilson class (gradients across the \linename\ are
measured in arbitrary units).

Figure \ref{fig:res:rlm}(a) is indicative of the other plots in this
figure, regardless of the transform used.  Over all \ls s, the
\bgdcl-class has larger average maximum gradients than the
\bgcl-class, which is larger than the \bcl-class, which is larger than
the \acl-class (excepting where low numbers of results mean that good
averages cannot be obtained).  The same ordering holds for all other
plots.  Hence when an active region moves from one class to another,
on average, the change in the geometrical structure of the active
region is reflected by a change in gradient content at all \ls s, and
not on any particular \ls.  All measures of the gradient given here
exhibit the same property - gradients in active region classes
regarded as being more likely to flare have larger maximum gradients,
larger average gradients, and larger standard deviations along the
\linename.  Figures \ref{fig:res:rlm}(c,f) suggest a greater
variability in the more active classes, and so suggest that more
active classes exhibit a wider range of gradient conditions, again, at
all \ls s.  This indicates that on average, the gradient content
between opposite polarity regions maintains the same Mt. Wilson
classification order, regardless of whether the field is understood as
being organized on small scales, or large scales.

\subsection{Weighted \ls\ versus weighted gradients}
\label{sec:weighted}
\begin{figure}
  \vspace*{0.05\textheight}
  \centerline{
    \hspace*{0.015\textwidth}
    \includegraphics[width=1.0\textwidth,height=0.40\textheight,clip=]{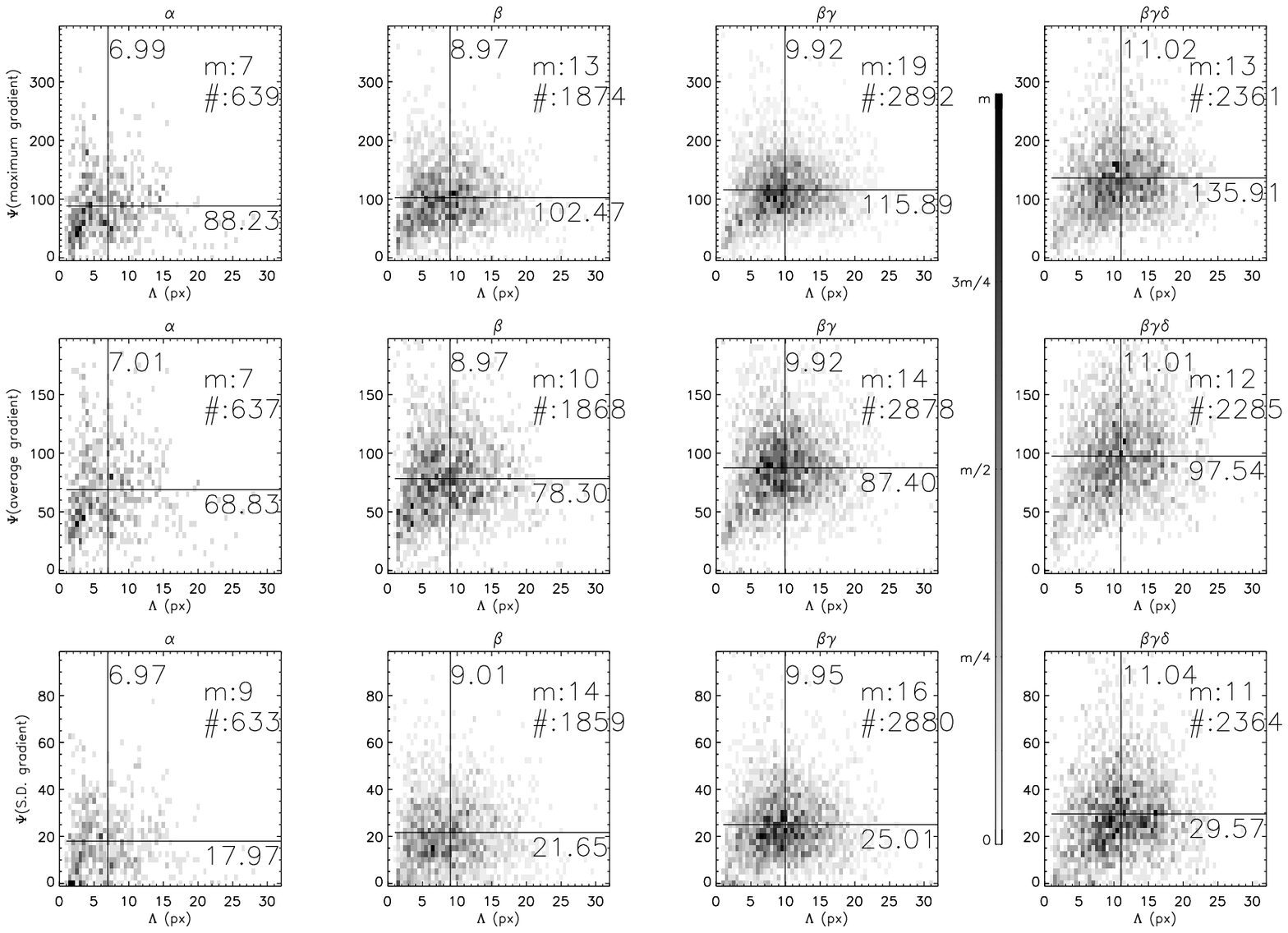}
  }    
  \vspace*{0.02\textheight}
  
  \centerline{
    \hspace*{0.000\textwidth}
    \includegraphics[width=1.0\textwidth,height=0.40\textheight,clip=]{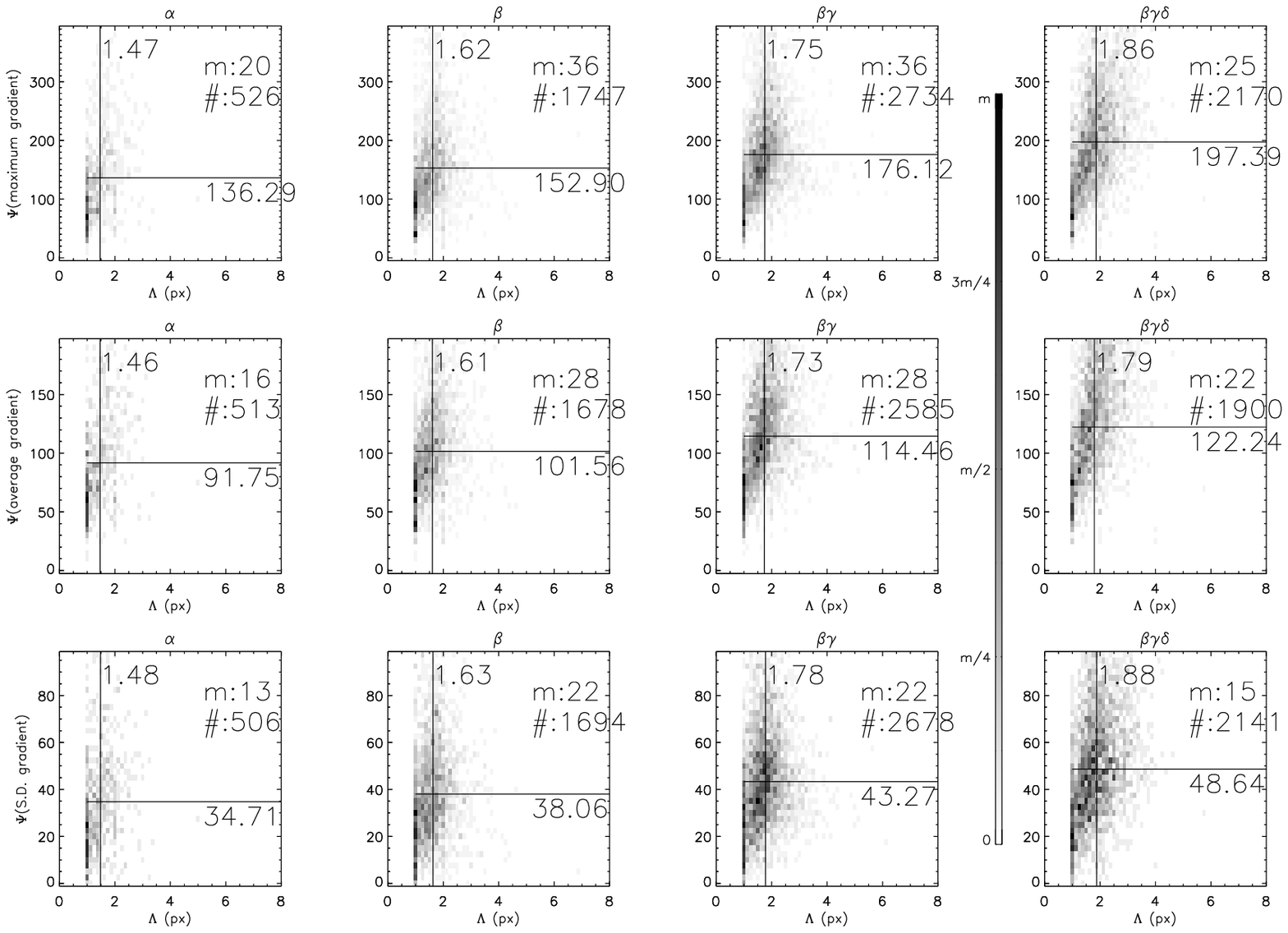}
  }
  \vspace{-1.38\textwidth}   % Shift close to the panel top 
  \centerline{\large \bf     % Includes the labels (here needs the color package)
    \hspace{0.0 \textwidth} \color{black}{(I): Mexican hat}
    \hfill}
  \vspace{0.655\textwidth}    % Shift back to the panel bottom 
  \centerline{\large \bf     % Includes the labels (here needs the color package)
    \hspace{0.0 \textwidth} \color{black}{(II): median transform}
    \hfill}
  \vspace{0.68\textwidth}    % Shift back to the panel bottom 
%   \vspace{-0.80\textwidth}   % Shift close to the panel top 
%   \centerline{\large \bf     % Includes the labels (here needs the color package)
%     \hspace{0.0 \textwidth} \color{white}{(I)}
%     \hfill}
%   \vspace{0.375\textwidth}    % Shift back to the panel bottom 
%   \centerline{\large \bf     % Includes the labels (here needs the color package)
%     \hspace{0.0 \textwidth} \color{white}{(II)}
%     \hfill}
%   \vspace{0.33\textwidth}    % Shift back to the panel bottom 
  \caption{Weighted average \ls\ ($\Lambda$) versus weighted average
    gradient statistic ($\Psi$(maximum gradient),$\Psi$(average
    gradient) and $\Psi$(standard deviation of gradient) ).
    Horizontal/vertical lines are annotated with the weighted average
    gradient/\ls\ of the distribution.  Also indicated on each plot is
    the maximum number of counts in a single histogram bin ('m') and
    the total number of counts in the distribution ('\#').  See
    Section \ref{sec:weighted} for more details on the calculation of
    weighted averages.}
  \label{fig:res:wav}
\end{figure}
The previous results show that active region fields may on average, be
ordered by Mt. Wilson classification without regard to any given \ls.
The above study also shows that gradients exist between objects of
many different \ls\ (active region fields are known to be
multi-fractal: \opencite{conlon}; \opencite{2005SoPh..228....5G};
\opencite{1996ApJ...465..425L}; \opencite{1993ApJ...417..805L} ).
However, a representative \ls s and gradient statistic can be defined
for a given active region, which summarizes the \mres\ nature of the
active region.  Assume that the \linename\ $q_{i}$ is found at \ls\
$L_{q_{i}}$ and has gradient statistic (either a maximum, average or
standard deviation of gradient) $G_{q_{i}}$.  Further, the local
wavelet power around $q_{i}$ is $w_{q_{i}}$.  If there are $n$
\linename s, $1\le i \le n$ for a given active region then the {\it
  weighted average \linename\ \ls} is
\begin{equation}
\label{eqn:Lambda}
\Lambda = \sum_{i=1}^{n}w_{q_{i}}{L_{q_{i}}}/\sum_{i=1}^{n}w_{q_{i}}
\end{equation}
and similarly, the {\it weighted average gradient statistic} is
\begin{equation}
\label{eqn:Psi}
\Psi = \sum_{i=1}^{n}w_{q_{i}}{G_{q_{i}}}/\sum_{i=1}^{n}w_{q_{i}}.
\end{equation}
Weighting by the local wavelet power (the same local wavelet power of
Section \ref{sec:algorithm}) takes into account the ``importance'' of the
\linename\ found at that location.

Figure \ref{fig:res:wav} plots frequency distributions of $\Lambda$
and $\Psi$ for both \mres\ analyses, for all 9757 magnetograms,
binned by Mt. Wilson classification.  Although the distributions are
highly scattered, it is clear that in all measures, and for both
\mres\ analyses, that average values of $\Lambda$ and $\Psi$
increase in the order
\acl$\rightarrow$\bcl$\rightarrow$\bgcl$\rightarrow$\bgdcl.  As the
Mt. Wilson classification increases in the above order, the
organization of the \linename\ moves to progressively longer \ls, and
the gradient on these separators also increases.  Hence there are
longer separators, with larger gradients, as the Mt. Wilson class
increases to those known to have a greater chance of activity.  The
order \acl$\rightarrow$\bcl$\rightarrow$\bgcl$\rightarrow$\bgdcl\ also
generally connotes an increase in active region size.  However, an
increasing size does not necessarily mean that there are longer
\linename\ between opposite polarity flux regions on those longer \ls;
it depends entirely on the locations and proximity of the opposite
polarity field as it emerges.  For a given size, a \bgdcl\ active
region has much more opposite polarity structure than other Mt. Wilson
classes (indeed, it is implicit in the definitions - see Table
\ref{tab:mtwilson}), and so the increase in $\Lambda$ and all three
$\Psi$ statistics (maximum gradient on a \linename, average gradient
on a \linename, and the standard deviation of the gradient on a
\linename), indicate the presence of increasingly diverse field
structure.  The qualitative result is the same for both the \mres\
results, indicating that both analyses are capturing consistent
behavior.

\subsection{Gradients in flaring and non-flaring active regions}
\label{sec:fnf}
\begin{figure}
  \vspace*{0.05\textheight}
  \centerline{
    \hspace*{0.015\textwidth}
    \includegraphics[width=1.0\textwidth,height=0.40\textheight,clip=]{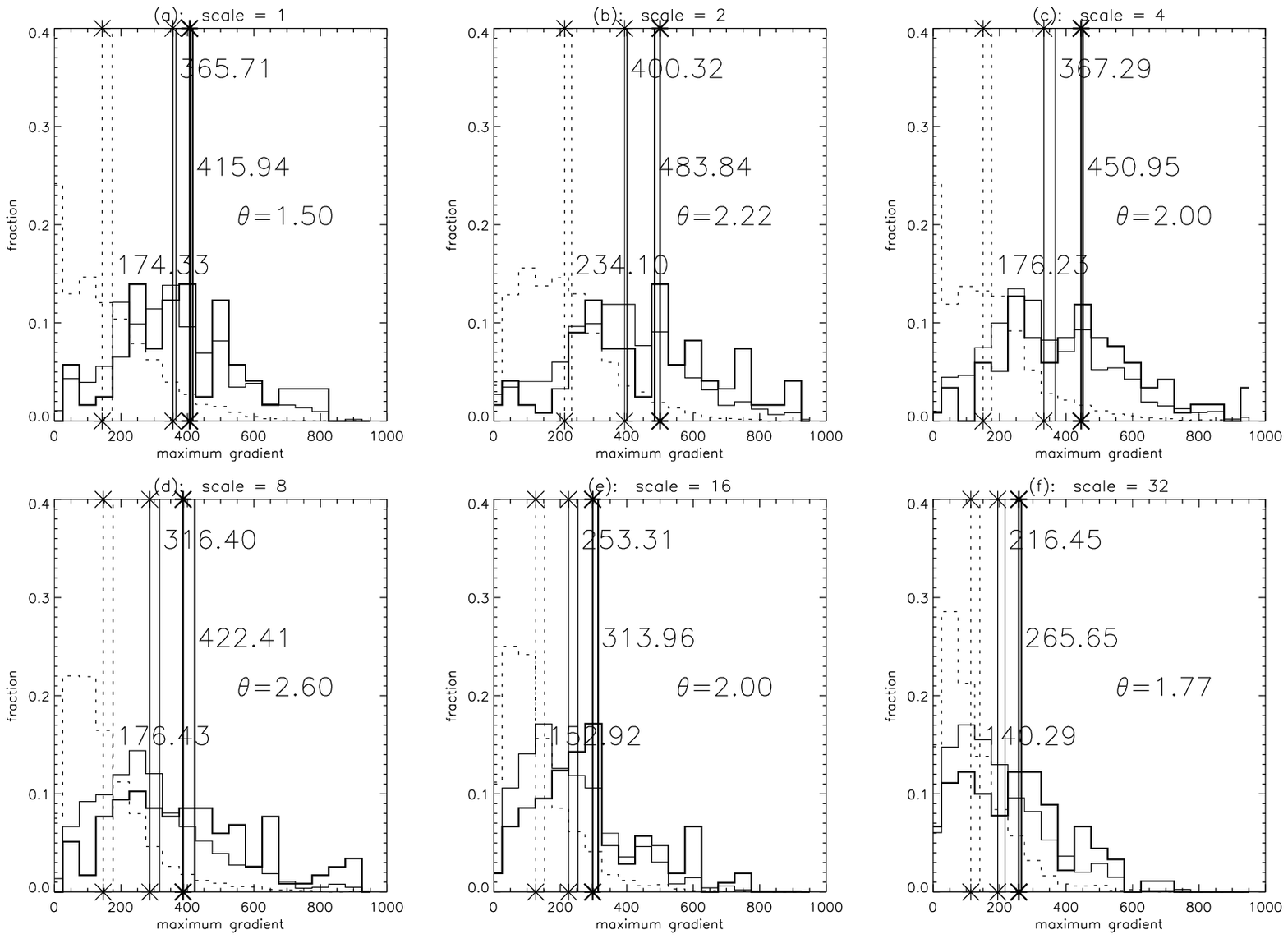}
  }    
  \vspace*{0.02\textheight}
  
  \centerline{
    \hspace*{0.000\textwidth}
    \includegraphics[width=1.0\textwidth,height=0.40\textheight,clip=]{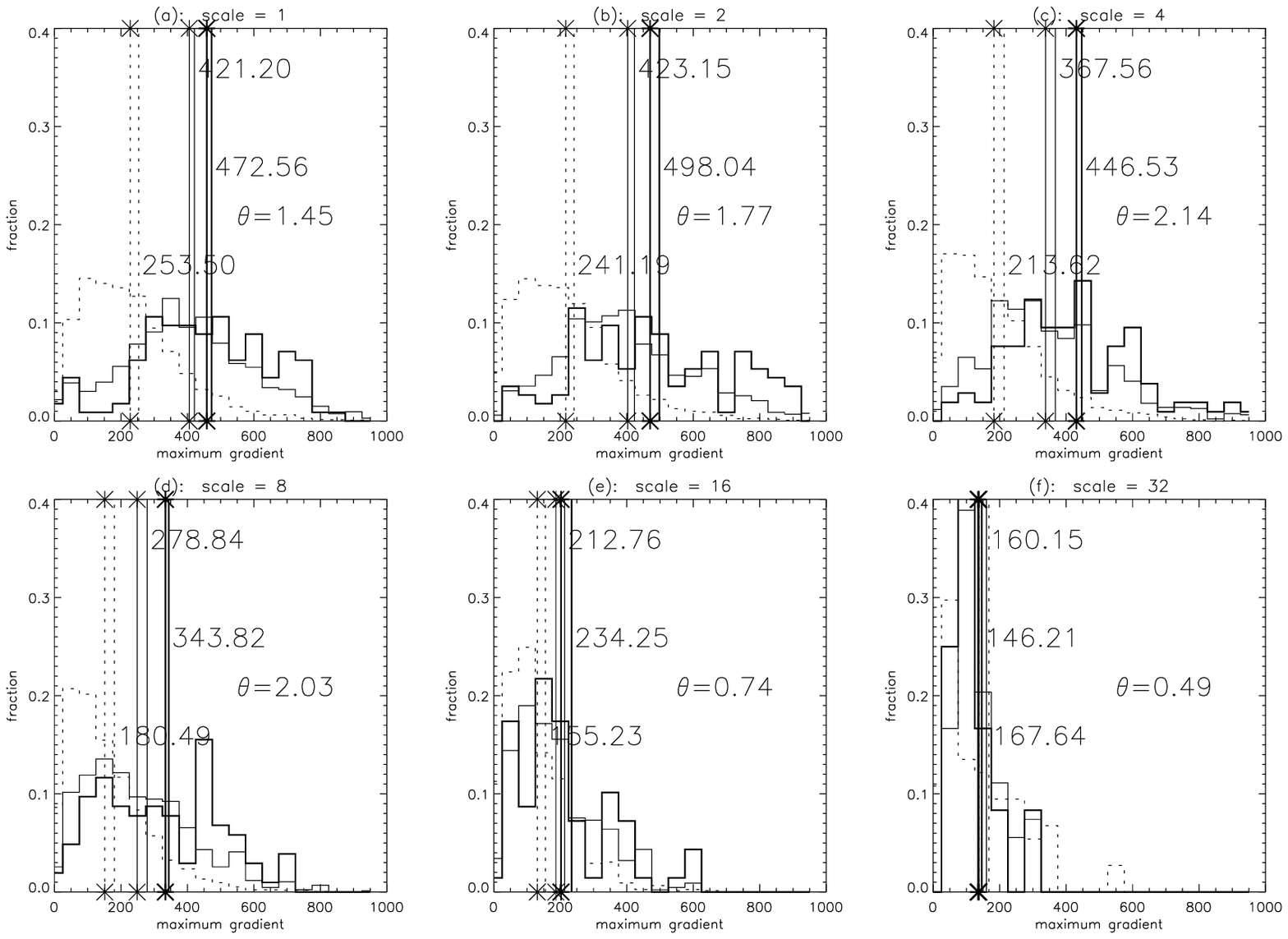}
  }
  \vspace{-1.38\textwidth}   % Shift close to the panel top 
  \centerline{\large \bf     % Includes the labels (here needs the color package)
    \hspace{0.0 \textwidth} \color{black}{(I): Mexican hat}
    \hfill}
  \vspace{0.655\textwidth}    % Shift back to the panel bottom 
  \centerline{\large \bf     % Includes the labels (here needs the color package)
    \hspace{0.0 \textwidth} \color{black}{(II): median transform}
    \hfill}
  \vspace{0.68\textwidth}    % Shift back to the panel bottom 
%   \vspace{-0.80\textwidth}   % Shift close to the panel top 
%   \centerline{\large \bf     % Includes the labels (here needs the color package)
%     \hspace{0.0 \textwidth} \color{white}{(I)}
%     \hfill}
%   \vspace{0.375\textwidth}    % Shift back to the panel bottom 
%   \centerline{\large \bf     % Includes the labels (here needs the color package)
%     \hspace{0.0 \textwidth} \color{white}{(II)}
%     \hfill}
%   \vspace{0.33\textwidth}    % Shift back to the panel bottom 
  \caption{Distributions of the maximum gradient found as a function
    of multi-resolution transform, analyzing gradient and flare
    occurence (only those flares that occur no later than six hours
    after the active region gradient measurement are included).  In
    all plots, the dotted lines refer to active regions having no
    flares occuring in them.  The thin solid line refers to active
    regions having at least one flare of class 'A' or above ({\it
      all-flare distribution}); the thick solid line refers to active
    regions having class 'M' or above ({\it large-flare
      distribution}).  Vertical lines without asterisks indicate the
    mean value of the gradient distribution; vertical lines with
    asterisks indicate the median value of the gradient distribution.
    The quantity $\theta$ is the test quantity calculated to apply the
    Kolmogorov-Smirnov test.  It is applied to test if the all-flare
    and large-flare distributions are different; values above 1.36
    confirm the null hypothesis (that the two distributions are drawn
    from the same underlying distribution) less than 5\% of the time -
    see Section \ref{sec:fnf} for more detail. }
  \label{fig:res:fnf}
\end{figure}

As suggested in the introduction, field gradients have long been
associated with flaring activity.  In this study, active regions are
split into three sets in order to quantify the relationship between
the presence of gradients between opposite polarity of different size
scales and the occurance and size of flares.  The first set of active
regions have no flaring activity associated with them at any time.
The second set of magnetograms (the {\it all-flare} set) have at least
one GOES 'A1.0' class flare, or more energetic, occuring no more than
six hours after the active region magnetogram.  The third set of
magnetograms (the {\it large flare} set ) have at least a cumulative
flare index equivalent to a GOES 'M1.0' class flare, or more
energetic, ocurring no more than six hours after the active region
magnetogram.  \linename\ for these magnetograms are calculated, and
distributions of the {\it maximum} gradients found are shown in Figure
\ref{fig:res:fnf}.

It is clear that, on average, the gradient content of a flaring active
region field is very different from that of a non-flaring active
region field.  Further, this difference is apparent on almost all \ls;
it does not matter which size scale one considers the opposite
polarity regions to be the one at which this difference is best
measured.  The difference between the all-flare and large-flare
distributions is less pronounced, but measurable both by looking at
the mean and median values of gradients, and also by the
Kolmogorov-Smirnov test.  The Kolmogorov-Smirnov test for two samples
(\opencite{wasserman}; \opencite{keeping}; \opencite{barlow}) treats
the agreement between two cumulative distributions\footnote{For a
  probability density function $f(x)$ the cumulative distribution
  function is $G(x) = \int_{0}^{x}f(y)dy$} $G_{1}$ and $G_{2}$ against
the null hypothesis that the two distributions are samples come from
the same underlying distribution function.  The first step is to
calculate
\begin{equation}
\label{eqn:KSA}
d = \max\left|G_{1}(x) - G_{2}(x)\right|.
\end{equation}
If $G_{1,2}$ has $n_{1,2}$ samples then it can be shown that 
\begin{equation}
\label{eqn:KSB}
\lim_{n_{1},n_{2}\rightarrow \infty} \mbox{prob}\left(
\theta \ge \lambda \right) = 2\sum_{m=1}^{\infty}(-1)^{m+1}e^{-2m^{2}\lambda^{2}}
\end{equation}
where $\theta = d\sqrt{N}$ and $N=n_{1}n_{2}/(n_{1} + n_{2})$.  Values
of $\theta \ge 1.36$ reject the null hypothesis at greater than or
equal to the 5\% level, that is, on average five times out of a
hundred the two cumulative distributions come from the same underlying
distribution function ($\lambda = 1.63$ is equivalent to 1\%,
$\lambda=1.73$, 0.5\% and $\lambda=1.95$, 0.1\%).  It is a
nonparametric test and is thus appropriate to use in this case where
the true gradient distribution is unknown.  

The Kolmogorov-Smirnov test shows that the Mexican hat derived
distributions for all-flare and large-flare maximum gradients, reject
the null hypothesis with a high degree of confidence, for all the \ls\
studied.  The median transform results also reject the null hypothesis
with a high degree of confidence, but only for the first four \ls s -
the hull hypothesis cannot be rejected at \ls = 16 and 32 pixels
(probably due to insufficient data at these \ls\ - see Sections
\ref{sec:median} and \ref{sec:compare}).  Hence the difference between
active regions that contain any type of flare, and active regions that
contain large flares is measureable in the active region gradient
content, without regard to which opposite polarity region \ls\ is
considered.

\section{Conclusions}
\label{sec:discuss}

Active region magnetic fields are known to exhibit multi-fractal
properties. The study presented here connects the geometrical
arrangement of opposite polarity regions (via the \linename\ concept)
to the \ls\ properties of the field.  There appears to be no special
\ls\ that determines the gradient content of the active region field,
which agrees with the notion from previous fractal and multi-fractal
studies that there is no preferred \ls\ to the spatial size of the
absolute valoe of active region flux elements.

The Mt. Wilson classification does capture some structure information
on the field, and through that, gradient content (Figures
\ref{fig:res:rlm} and \ref{fig:res:wav}).  The Mt. Wilson
classification is based on the very largest \ls\ describing the
geometrical arrangement of the field.  Figure \ref{fig:res:rlm} shows
that regardless of the \ls\ used, the gradient content of different
classes of active region field can be distinguished on average.  This
is a result of the multi-scale nature of the active region magnetic
field.  Figure \ref{fig:res:wav} shows that even although all
Mt. Wilson classes retain their multi-scale structure (i.e., gradients
at all \ls\ are present), there is a general increase in the average
\ls\ and gradient as a function of Mt. Wilson class, in the order
\acl$\rightarrow$\bcl$\rightarrow$\bgcl$\rightarrow$\bgdcl).  Not only
is the active region field appearing at larger \ls, the organization
of its gradients is also appearing at larger \ls, and those gradients
are getting larger.

The equivalence of all \ls s in the active region magnetic field is
also shown in Figure \ref{fig:res:fnf}; at all \ls, there are
measurable differences in the gradient content for active regions that
give rise to large flares, active regions that give rise to all sizes
of flares, and active regions that do not have flares.  On average,
one can distinguish the difference between these types of active
region by considering the geometrical arrangement of the opposite
polarity flux at any \ls.  The geometrical arrangement of opposite
polarity field preserves the multi-scale nature of the active region
at all \ls, through to the gradient content of the field.

When an active region breaks through the photosphere, injection
happens at multiple \ls s.  The largest gradients between opposite
polarity regions are observed at the smallest \ls\ (see Figure
\ref{fig:res:rlm}(a,d)).  Although the largest gradients may be at the
smallest \ls s, gradients also exist between opposite polarity regions
of longer \ls.  This study suggests that indicators of flare activity
and Mt. Wilson classification exist at all \ls s, and that all active
regions \ls\ are equivalent, due to the multi-scale properties of the
field.

\begin{acks}
  This work was supported by NASA Living With a Star Targeted Research
  and Technology award NNH04CC65C. SOHO is a joint project of
  international co-operation by ESA and NASA.
\end{acks}

% format of references provided by the journal (.bst)
\bibliographystyle{spr-mp-sola}

% name your Bibtex file containing your references (.bib)
\bibliography{ip}

\end{article} 
\end{document}